\input iopppt.modifie
\input epsf.tex
\def\received#1{\insertspace 
     \parindent=\secindent\ifppt\textfonts\else\smallfonts\fi 
     \hang{Received #1}\rm } 
\headline={\ifodd\pageno{\ifnum\pageno=\firstpage\titlehead
   \else\rrhead\fi}\else\lrhead\fi} 

\def\rrhead{\textfonts\hskip\secindent\it 
    \shorttitle\hfill\rm L\folio} 
\def\lrhead{\textfonts\hbox to\secindent{\rm L\folio\hss}%
   \it\aunames\hss} 
\footline={\ifnum\pageno=\firstpage
\hfil\textfonts\rm L\folio\fi}   
\def\titlehead{\smallfonts J. Phys. A: Math. Gen. {\bf 18} (1985)
L325--L329 
\hfil} 

\firstpage=325
\pageno=325


\jnlstyle

\jl{1}

\letter{Conformal invariance and linear defects in the\break
two-dimensional Ising model}

\author{L Turban}

\address{Laboratoire de Physique du
Solide\footnote{$\dagger$}{Laboratoire associ\'e au CNRS No 155.},
ENSMIM, Parc de Saurupt, F 54042 Nancy Cedex, France}

\received{5 February 1985}

\abs
{Using conformal invariance, we show that the non-universal
exponent  $\eta_0$ associated with the decay of correlations along a
defect line of modified bonds in the square-lattice Ising model is
related to the amplitude $A_0\!=\!\xi_n/n$ of the correlation length
$\xi_n(K_{\rm c})$ at the bulk critical coupling $K_{\rm c}$, on a
strip with width $n$, periodic boundary conditions and two
equidistant defect lines along the strip, through
$A_0=(\pi\eta_0)^{-1}$.
\endabs

\vglue1cm

In addition to scale invariance, a statistical system at its
critical point also displays conformal invariance (Polyakov 1970,
Fisher 1973, Wegner 1976). This property may be used to constrain
the form of the correlation functions. In two-dimensional systems it
even completely determines the critical exponents and correlation
functions in the bulk (Belavin \etal 1984, Dotsenko 1984, Friedan
\etal 1984) or near a surface (Cardy 1984a).

Recently conformal invariance has also been used (Cardy 1984b) to
justify a remarkable universal relation between the amplitude $A_0$
of the correlation length $\xi_n$ on a strip with periodic boundary
conditions and the bulk critical exponent $eta$ for large $n$ values:
$$
\eqalignno{
\xi_n&=A_0n&(1)\cr
A_0&=(\pi\eta)^{-1}\,.&(2)\cr}
$$
This relation was first established in the case of Anderson
localisation (Pichard and Sarma 1981); it was verified on the 2D $XY$
model (Luck 1982), the Potts model (Derrida and de S\`eze 1982),
generalised for correlations other than of the order-order type and
anisotropic systems (Nightingale and Bl\"ote 1983) and for quantum
systems (Penson and Kolb 1984).

The purpose of the present letter is to establish a similar
relation between the correlation length amplitude $A_0$ in the
strip geometry and the non-universal exponent $\eta_0$ which
characterises the decay of the order parameter correlation function
at the critical point near a defect line in the 2D Ising model
through a conformal transformation.

{\par\begingroup\medskip
\epsfxsize=7cm
\topinsert
\centerline{\epsfbox{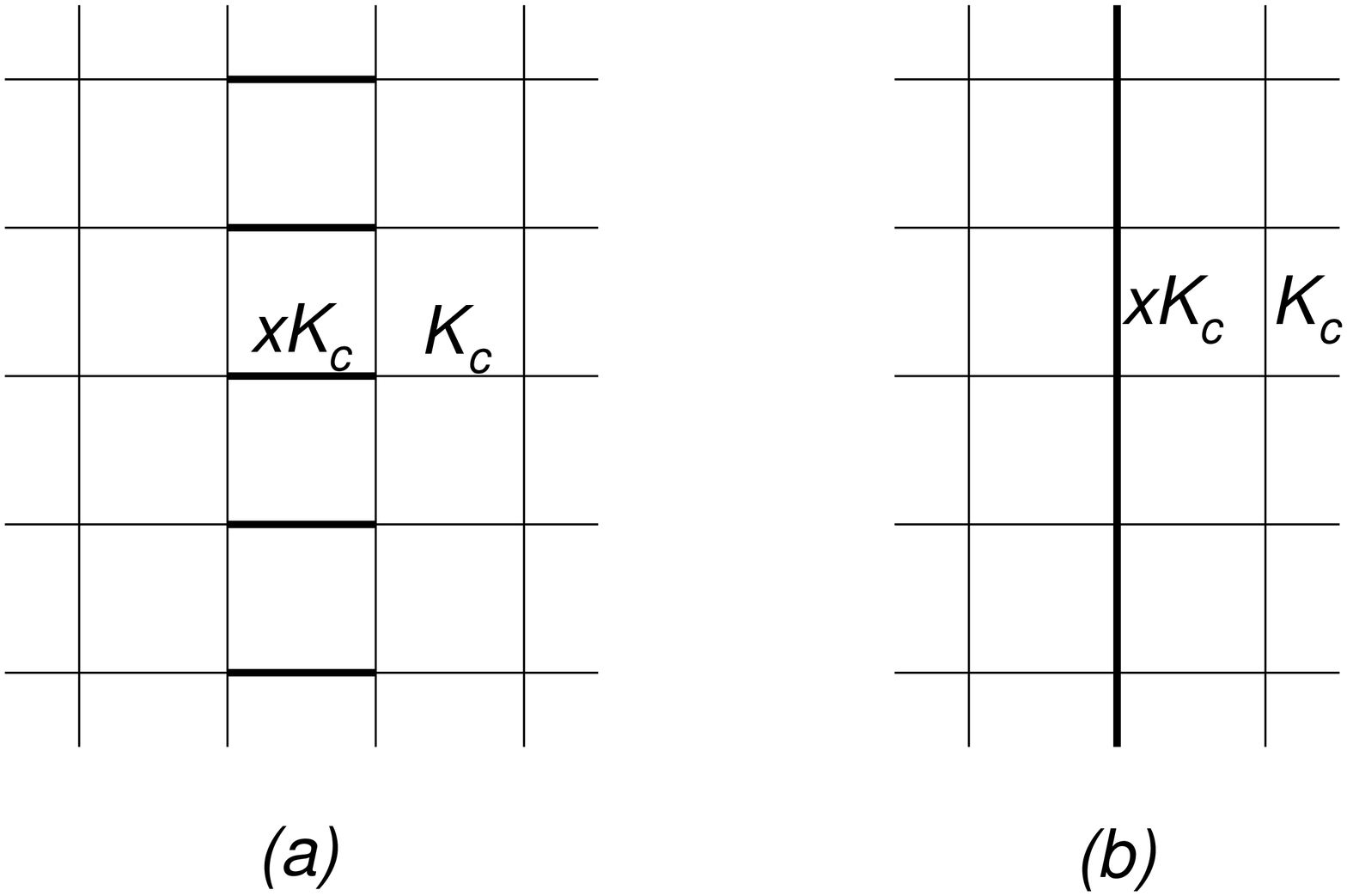}}
\smallskip
\figure{Defect line in: (a) the ladder geometry and (b) the chain
geometry along which the interaction strength is modified ($K'=xK$,
where $K$ is the bulk interaction strength).} 
\endinsert
\endgroup
\par}

We shall consider two types of defect lines, the ladder and chain
cases where the interaction strength $K$ is modofied along a straight
line on bonds which are either perpendicular or parallel to the line
(figure 1). Exact results have been obtained showing that for the
Ising model the defect exponent $\eta_0$ is non-universal (Bariev
1980, McCoy and Perk 1980). For two points at a fixed distance from
the defect, the correlation function at the critical point of the
bulk decays like 
$$
\langle\varphi(\bi{r}_1)\varphi(\bi{r}_2)\rangle\sim r^{-\eta_0}
\qquad
(r=\vert\bi{r}_1-\bi{r}_2\vert)
\eqno(3)
$$
where $\varphi(\bi{r})$ is the Ising scalar field. The dependence on
the defect strength for the square lattice is as follows
$$
\eta_0=\left[\frac{1}{\pi}\cos^{-1}(\chi)\right]^2
\eqno(4)
$$
where
$$
\chi={\cosh(2xK_{\rm c})-\cosh(2K_{\rm c})\over\cosh(2K_{\rm c})
\cosh(2xK_{\rm c})-1}
\eqno(5)
$$
in the ladder geometry and
$$
\chi=\tanh[2K_{\rm c}(x-1)]
\eqno(6)
$$
in the chain geometry. $K_{\rm c}\!=\!\frac{1}{2}\ln(1\!+\!\sqrt{2})$
is the Ising square lattice critical coupling.

{\par\begingroup\medskip
\epsfxsize=7cm
\topinsert
\centerline{\epsfbox{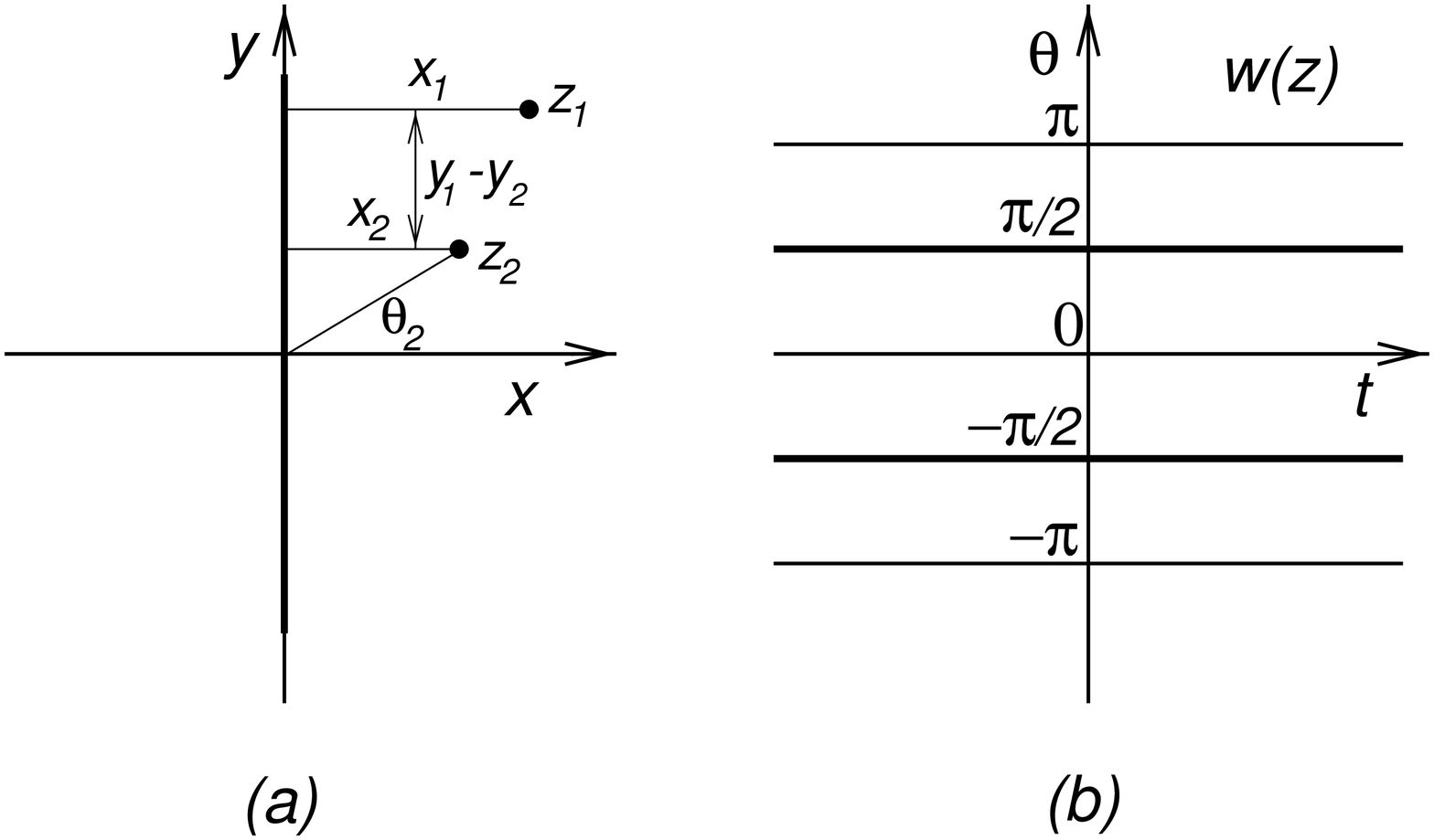}}
\smallskip
\figure{The conformal transformation $w\!=\!\ln(z)$ maps the plane
with a defect line (a) into a strip with width $2\pi$ and two defect
lines (b).}  
\endinsert
\endgroup
\par}
 
Following Cardy (1984b) we use a finite conformal
transformation 
$$ 
w(z)=\ln(z)
\eqno(7)
$$
to map the plane with a linear defect into a strip $\vert{\rm
Im}(w)\vert\!\lequal\!\pi$ with {\it two linear defects} along the
strip at $\vert{\rm Im}(w)\vert\!=\!\pi/2$ and periodic boundary
conditions (figure~2). Conformal covariance of the correlation
function leads to
$$
\langle\varphi(z_1)\varphi(z_2)\rangle=\vert w'(z_1)\vert^x
\vert w'(z_2)\vert^x\langle\varphi(w_1)\varphi(w_2)\rangle
\eqno(8)
$$
where $x\!=\!\eta/2$ is the anomalous dimension of the field
$\varphi$. Writing $z_j\!=\!\exp(t_j\!+\!\i\theta_j)$, we obtain
$$ 
\langle\varphi(t_1+\i\theta_1)\varphi(t_2+\i\theta_2)\rangle_{\rm
s}=\exp[x(t_1+t_2)]\langle\varphi(z_1)\varphi(z_2)\rangle 
\eqno(9)
$$
where the correlation function on the left-hand side is evaluated in
the strip geometry whereas on the right-hand side it is taken on the
infinite plane. Through an infinitesimal conformal transformation,
one may show that in the infinite plane with a defect line, like in
the case of a surface (Cardy 1984a), the correlation function
behaves as follows
$$
\langle\varphi(z_1)\varphi(z_2)\rangle=\vert x_1x_2\vert^{-x}
\Phi\left[{(y_1-y_2)^2+x_1^2+x_2^2\over x_1x_2}\right]
\eqno(10)
$$
where $z_j\!=\!x_j\!+\!y_j$. The asymptotic behaviour near the
defect requires that $\Phi(\rho)$ behaves as $\rho^{-x_0}$ where
$x_0$ is a defect exponent ($x_0\!=\!\eta_0/2$) as its argument
$\rho\!\rightarrow\!\infty$. Changing for the strip variables, one
obtains
$$
\eqalign{
\langle\varphi(z_1)\varphi(z_2)\rangle=&\exp[-x(t_1+t_2)]  
\vert\cos\theta_1\cos\theta_2\vert^{-x}\cr
&\times\Phi\left[{\exp(t_1-t_2)+\exp(t_2-t_1)
-2\sin\theta_1\sin\theta_2\over\cos\theta_1\cos\theta_2}\right]\cr}
\eqno(11)
$$
and in the limit where $t_1-t_2\!\rightarrow\!+\!\infty$
$$
\langle\varphi(z_1)\varphi(z_2)\rangle\sim
\exp[-x(t_1+t_2)-x_0(t_1-t_2)] 
\vert\cos\theta_1\cos\theta_2\vert^{x_0-x}
\eqno(12)
$$
so that on the strip
$$
\langle\varphi(t_1+\i\theta_1)\varphi(t_2+\i\theta_2)\rangle_{\rm s}
\sim f(\theta_1,\theta_2)\exp[-x_0(t_1-t_2)]
\eqno(13)
$$
and the correlations along the strip with two defect lines decrease
exponentially with a correlation length
$$
\xi={1\over x_0}={2\over\eta_0}
\eqno(14)
$$
measured in units where the width of the strio is $2\pi$. If one
introduces a lattice, $n$ lattice spacings wide, the correlation
length $\xi_n$, measured in lattice spacing units, satisfies:
$$
\eqalignno{
&\xi_n\,{2\pi\over n}={2\over\eta_0}&(15)\cr
&\xi_n={n\over\pi\eta_0}&(16)\cr}
$$
so that the universal relation between the correlation length
amplitude and the decay exponent remains true if one replaces the
bulk exponent $\eta$ by the defect exponent $\eta_0$ and works on a
strip with two defect lines.

\par\topinsert\fl\hglue-.58cm
\vbox{\hsize=8.33cm\parindent=0pt
\epsfxsize=5.8cm
\centerline{\hglue1.67cm\epsfbox{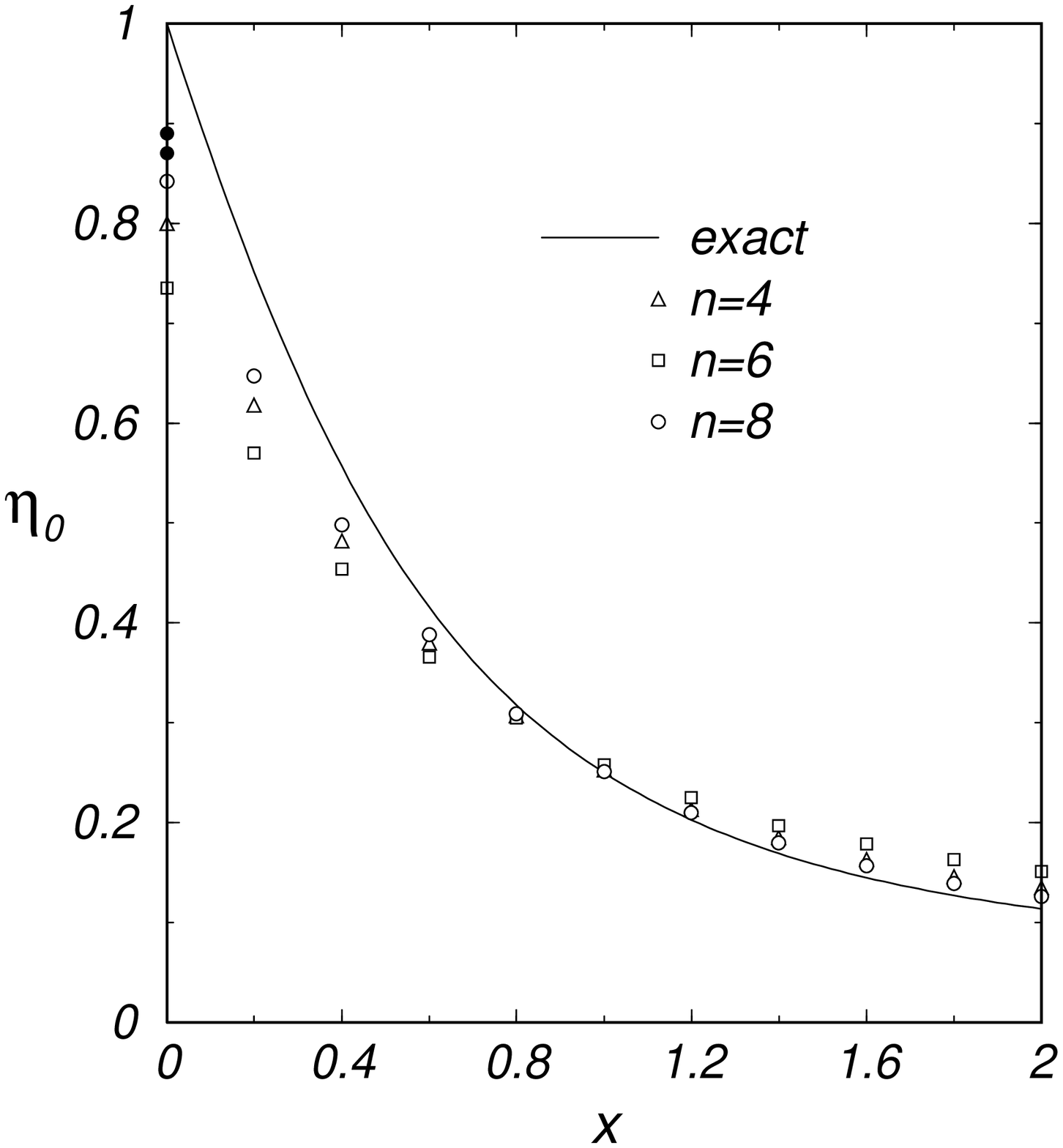}}
}
\hglue-1.67cm\vbox{\hsize=8.33cm\parindent=0pt
\epsfxsize=5.8cm
\centerline{\hglue1.67cm\epsfbox{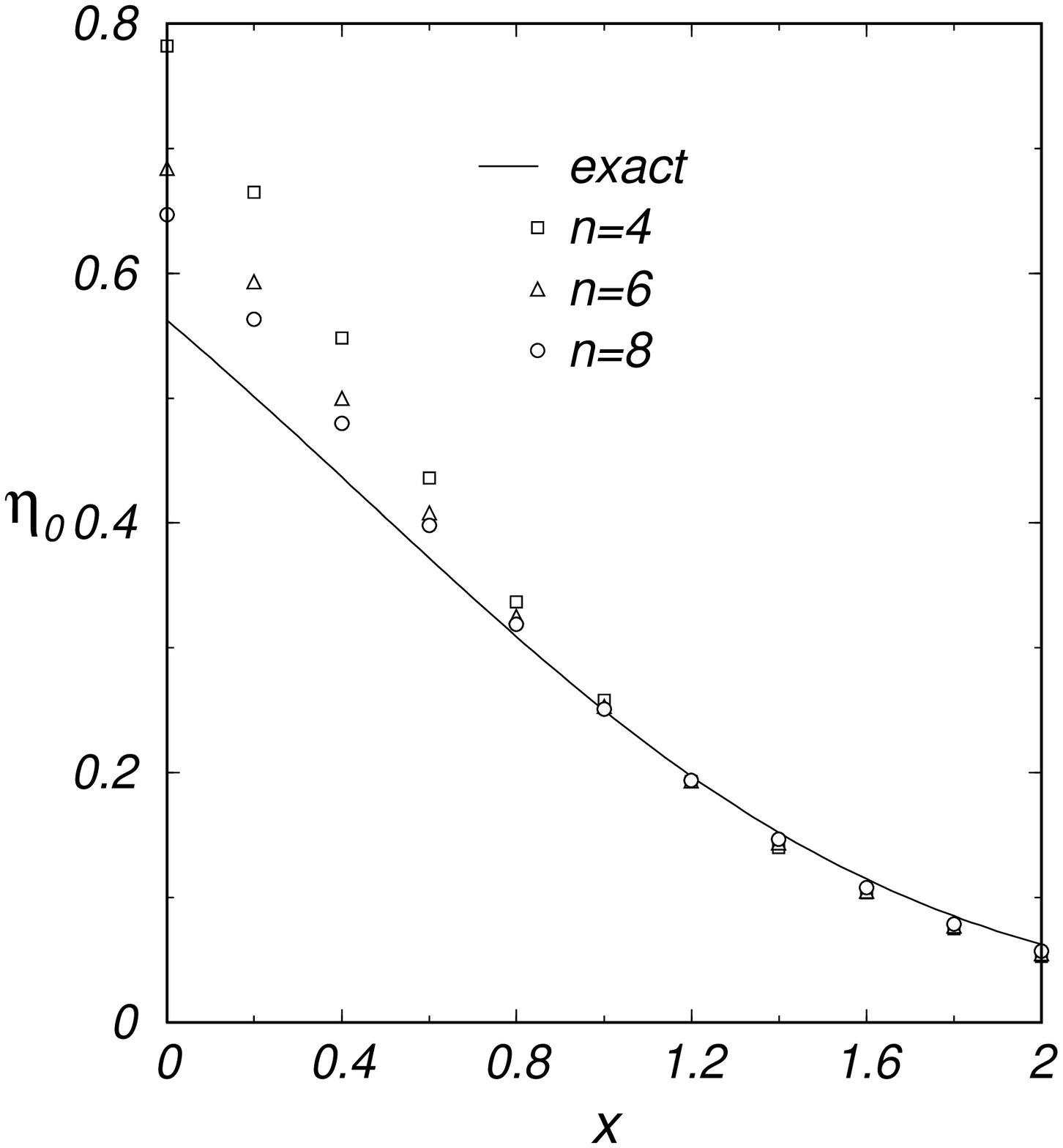}}
}
\smallskip
\figure{Defect exponent $\eta_0$ as a function of $x=K'/K_{\rm c}$
where $K'$ is the modified interaction along the defect line in
the ladder (left) and chain (right) geometries. Approximate
results are deduced from the correlation length amplitude
$A_0=\xi_n/n=(\pi\eta_0)^{-1}$ on strips with width $n=4,6,8$. In
the ladder geometry results are also given for $n=10,12$ when
$x=0$ (full circles). The convergence is slow on the side where
the interaction strength is weaker than in the bulk.}   
\endinsert
\par 
 
The universal relation has been tested numerically using the
transfer matrix method on strips with increasing size ($n=4,6,8$)
and varying the defect strength ($x=0$ to $2$) either in the ladder
geometry or in the chain geometry. The results are given in
figure~3. The agreement with equation~(16) is satisfactory although
the convergence towards the exact curve on the side where the defect
interaction is weaker than in the the bulk is rather slow. In order
to test this convergence, we have extended the calculation up to
$n\!=\!12$ in the limit $x\!=\!0$ in the ladder geometry. Then the
system decouples into two strips with free boundary conditions and
with half the width of the initial strip with periodic boundary
conditions so that one may work on narrower strips to obtain
equivalent results. Clearly the convergence is quite slow there.

To conclude let us mention that various defect geometries may be
studied using the strip method. For instance a half-infinite defect
line leads to a strip with a single defect line and periodic
boundary conditions wheras a defect line perpendicular to a free
surface corresponds to a strip with a single defect line and free
boundary conditions.

\references
\refjl{Bariev R Z 1980}{Sov. Phys.-JETP}{50}{613}
\refjl{Belavin A A, Polyakov A M and Zamolodchikov A B 1984}{J.
Stat. Phys.}{34}{763}
\refjl{Cardy J L 1984a}{\NP\ {\rm B}}{240}{514}
\refjl{\dash 1984b}{\JPA}{17}{L385}
\refjl{Derrida B and de S\`eze J 1982}{J. Physique}{43}{475}
\refjl{Dotsenko V S 1984}{\NP\ {\rm B}}{235}{54}
\refbk{Fisher M E 1973}{\it Proc. 24th Nobel Symposium on Collective
Properties of Physical Systems}{ed B~Lundqvist and S~Lundqvist
(New-york: Academic) p 16}
\refjl{Friedan D, Qiu Z and Shenker S 1984}{\PRL}{52}{1575}
\refjl{Luck J M 1982}{\JPA}{15}{L169}
\refjl{McCoy B M and Perk J H H 1980}{\PRL}{44}{840}
\refjl{Nightingale M P and Bl\"ote H W J 1983}{\JPA}{16}{L657}
\refjl{Penson K A and Kolb M 1984}{\PR\ {\rm B}}{29}{2857}
\refjl{Pichard J L and Sarma G 1981}{J. Phys. C: Solid State Phys.}{14}{L617}
\refjl{Polyakov A M 1970}{JETP Lett.}{12}{381}
\refbk{Wegner F J 1976}{Phase Transitions and Critical
Phenomena}{vol 6, ed C Domb and M S Green (London: Academic) p 7}

\vfill\eject\bye